\documentclass[12pt]{article}

\usepackage{epsfig}			%standard package. Note graphics<graphicx<epsfig 

\newcommand{\nb}[0]{{N_{B}}}
\newcommand{\ns}[0]{{N_{S}}}

\newcommand{\sx}[0]{\sigma_x}
\newcommand{\sy}[0]{\sigma_y}

\newcommand{\cnotyes}[2]{\sx(#2)^{n(#1)}}

\newcommand{\vb}[0]{{\vec{b}}}

\newcommand{\beq}{\begin{equation}}  
\newcommand{\eeq}{\end{equation}}
\newcommand{\beqa}{\begin{eqnarray*}}  
\newcommand{\eeqa}{\end{eqnarray*}}

\newcommand{\enote}[1]{\cite{#1}}    %call it \enote (endnote) rather than \cite
\newcommand{\eqlabel}[1]{\renewcommand{\theequation}{#1}}%this allows me to label 
\begin{document}
\title{An Optimization for Qubiter}

\author{Robert R. Tucci\\
        P.O. Box 226\\ 
        Bedford,  MA   01730\\
        tucci@ar-tiste.com}

\date{ \today} 

\maketitle

\vskip2cm
\section*{Abstract}
In a previous paper, we described
a computer program called Qubiter which can decompose
an arbitrary unitary matrix into elementary operations
of the type used in quantum computation. In this paper,
we describe a method of reducing the number of elementary
operations in such decompositions.

\newpage

According to  Eqs.(4.17) and (4.18) of Ref.\enote{Tucci98a},
if we define $\Gamma(\theta)$ by

\beq
\Gamma(\theta) = 
\exp( i \theta \sy )
\otimes I_2 \otimes I_2
\;,
\eqlabel{1}\eeq
then

\beq
D = \prod_{\vb \in Bool^2} A_{\vb}
\;,
\eqlabel{2}\eeq
where

\beq
A_{00} = \Gamma(\theta_{00})
\;,
\eqlabel{3a}\eeq

\beq
A_{10} = 
\cnotyes{1}{2}\odot
\Gamma(\theta_{10})
\;,
\eqlabel{3b}\eeq

\beq
A_{11} = 
[\cnotyes{1}{2}
\cnotyes{0}{2}]
\odot
\Gamma(\theta_{11})
\;,
\eqlabel{3c}\eeq

\beq
A_{01} = 
\cnotyes{0}{2}\odot
\Gamma(\theta_{01})
\;.
\eqlabel{3d}\eeq
The operators $A_\vb$ commute so Eq.(2) is valid regardless of
the order in which the $A_\vb$ are multiplied.
Suppose we multiply them in the following order:

\beq
D = A_{01} A_{11} A_{10} A_{00}
\;.
\eqlabel{4}\eeq
Note that in this order, if $A_{\vb_L}$ is immediately to the left of $A_{\vb_R}$, then 
$\vb_L$ and $\vb_R$ are elements of $Bool^2$ that differ only in the
value of one bit. Eq.(4) simplifies to 

\beq
D = 
\cnotyes{0}{2}\Gamma(\theta_{01})
\cnotyes{1}{2}\Gamma(\theta_{11})
\cnotyes{0}{2}\Gamma(\theta_{10})
\cnotyes{1}{2}\Gamma(\theta_{00})
\;.
\eqlabel{5}\eeq
We see that by ordering the $A_\vb$ in this special way,
many c-nots cancel out. Only one c-not remains between any two adjacent 
$\Gamma$'s. There is no c-not to the right of the rightmost $\Gamma$ in Eq.(5).
And, because the leftmost $\vb$ in Eq.(4)
has only one non-zero bit, there is only one c-not to the left of the leftmost
$\Gamma$ in Eq.(5).

The above example
assumes $\nb=3$ and that we are decomposing a central matrix of type 1 
(i.e., a central matrix which is a single D matrix). 
However, this method of reducing the number of c-nots
in a decomposition can also be used for other values of $\nb$ and for the other two  
types of central matrices. Let's see how.
For general $\nb$, one can (see Appendix) order the elements $\vb$ of $Bool^{\nb}$
so that adjacent $\vb$'s differ only in the value of one bit, and so that the first $\vb$
is the zero vector and the last $\vb$ is a standard unit vector (i.e., it has only one non-zero bit).
We'll call such an ordering of the elements of $Bool^{\nb}$ a ``lazy" ordering
(a step from one $\vb$ to the next is ``lazy" since it involves a single bit flip instead of many).
Products of $A_\vb$ operators arise in the decomposition of all three types of central matrices.
If the order in which one multiples the $A_\vb$ is such that their $\vb$ subscripts are in a lazy order,
then the number of c-nots in the decomposition will be reduced.

\section*{APPENDIX: A Method of Generating a Lazy Ordering of $Bool^\nb$}

Create a binary tree $T$ with $\nb-1$ levels. Label each node of $T$
by its level number (top is 0, bottom is $\nb-1$). See Fig.1 for $T$ when $\nb = 3$.
Create a sequence $s = (s_1, s_2,  \cdots, s_{\ns-1}) $ of integers by recording the node labels
encountered during an  ``inorder transversal'' of the tree. For example,  for Fig.1, $s= (2,1,2,0,2,1,2)$.
Let $\vec{b}_0 = \vec{0} \in Bool^{\nb}$. On step $j\in \{1, 2, \ldots, \ns-1\}$, $\vec{b}_{j-1}$
 transforms into $\vec{b}_{j}$ 
when you flip the bit at position $s_j$. As usual, we label the bit positions so that they increase
from right to left and the rightmost bit is at position 0.

%don't float in general
%		\begin{figure}
		\begin{center}
			\epsfig{file=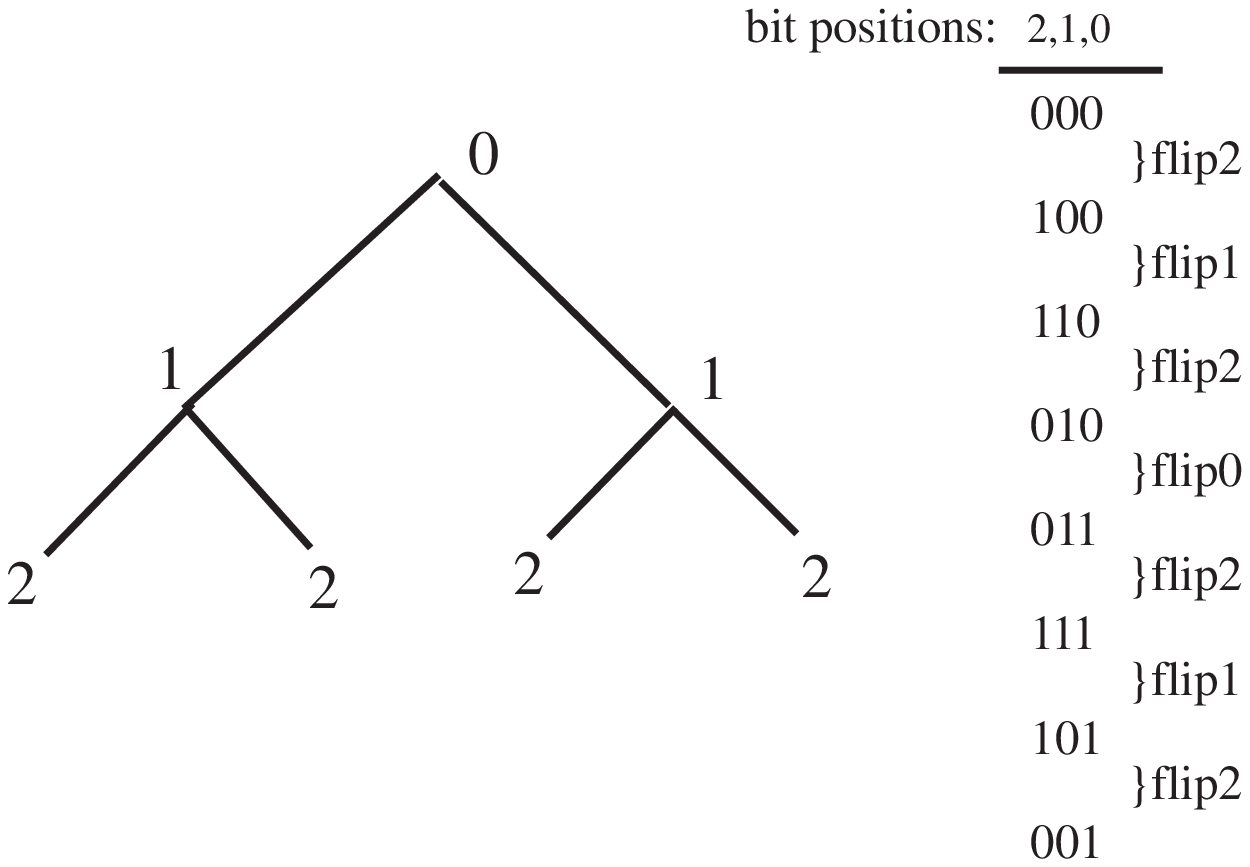}
			%\caption
			
			{Fig.1 A lazy ordering of $Bool^3$}
			%\label{}
		\end{center}
%		\end{figure}

\end{document}